# Unveiling the Skills and Responsibilities of Serverless Practitioners: An Empirical Investigation

Muhammad Hamza, Vy Kauppinen, Muhammad Azeem Akbar, Wardah Naeem Awan, Kari Smolander
Software Engineering Department, Lappeenranta-Lahti University of Technology,
15210 Lappeenranta, Finland
muhammad.hamza, azeem.akbar, vy kauppinen, wardah.awan,
kari.smolander@lut.fi

**Abstract**

*Enterprises are increasingly adopting serverless computing to enhance scalability, reduce costs, and improve efficiency. However, this shift introduces new responsibilities and necessitates a distinct set of skills for practitioners. This study aims to identify and organize the industry requirements for serverless practitioners by conducting a qualitative analysis of 141 job advertisements from seven countries. We developed comprehensive taxonomies of roles, responsibilities, and skills, categorizing 19 responsibilities into four themes: software development, infrastructure and operations, professional development and leadership, and software business. Additionally, we identified 28 hard skills mapped into seven themes and 32 soft skills mapped into eight themes, with the six most demanded soft skills being communication proficiency, continuous learning and adaptability, collaborative teamwork, problem-solving and analytical skills, leadership excellence, and project management. Our findings contribute to understanding the organizational structures and training requirements for effective serverless computing adoption.*

## 1. Introduction

Serverless computing revolutionizing the way software is developed and deployed, offering a model where developers can focus only on writing code without worrying about the underlying infrastructure. This paradigm shift allows enhanced scalability, reduced operational costs, and improved efficiency, as users only pay for the exact resources consumed during execution (Eivy & Weinman, 2017). Enterprises adopt serverless architecture to streamline the complexity of their software systems (Hamza et al., 2024). Major companies like Amazon Prime Video, Netflix, and Coca-Cola leverage serverless to improve scalability and efficiency in their operations (Baldini et al., 2017a).

However, the adoption of serverless computing introduces new responsibilities for practitioners, necessitating a distinct set of skills to tackle the challenges and maximize the benefits of this architecture. Practitioners and organizations often struggle to adopt serverless architecture due to a lack of relevant skills and understanding of crucial non-functional aspects. Current literature primarily focuses on the technical aspects of serverless computing, such as performance optimization, scalability, and cost efficiency (Baldini et al., 2017b). However, Serverless practitioners are required to be proficient in cloud platforms such as AWS Lambda, Azure Functions, Google Cloud Functions, and event-driven programming (Hamza et al., 2024).

Existing research outlines the skills needed by practitioners in other software engineering domains, such as requirements engineering (Daneva et al., 2017), testing (Cerioli et al., 2020), Microservices-based architecture (Ali et al., 2020), AI engineering (Meesters et al., 2022), and microservices architecture (Ayas et al., 2024). Nevertheless, there is limited empirical evidence on the specific requirements required from the practitioners to thrive in developing decoupled, event-driven, and stateless applications using serverless computing (Hamza et al., 2024).

This study aims to provide a comprehensive understanding of the technical roles, responsibilities, and skills necessary for developing serverless-based applications. To meet this objective, we perform an extensive qualitative analysis of job ads from Glassdoor. We specifically examine 141 job ads, selected from a dataset of 2,508 serverless-related job ads from seven English-speaking countries. Further, the analysis utilizes thematic analysis techniques to address the following research questions. The study aims to bridge the gap between industry demands and practitioner capabilities. Our findings contribute to understanding the organizational structures and training requirements for effective serverless computing adoption.

*RQ1: What are the roles and responsibilities of serverless practitioners as identified in job ads?*
*RQ2: What hard and soft skills are required from the serverless practitioner as identified in job ads?*
*RQ3: What are the trends in programming languages, serverless technologies, and cloud platforms as indicated in serverless- job ads?*

The rest of the paper is organized as follows: Section 2 covers related work, Section 3 outlines the methodology, Section 4 presents results, Section 6 provides discussion, and Section 7 concludes the study.

## 2. Related work

Organizations and researchers are increasingly focused on identifying the suitable engineering profiles required for technical capabilities in recruitment, training, and talent retention (Brooks, 2021). Recent studies have expanded this focus to understand the specific requirements organizations need from engineers in the evolving landscape of technology (Montandon et al., 2021). The unique nature of serverless architecture has prompted researchers and practitioners to explore various aspects, including architectural design, performance improvement, technological advancements, and challenges in testing and debugging (Hamza, 2023).

For instance, (Wen et al., 2021) conducted an in-depth analysis of 619 discussions from Stack Overflow, identifying the challenges developers face when developing serverless applications. Similarly, (Eskandani & Salvaneschi, 2023) analyzed 2,000 real-world open-source applications on GitHub, providing insights into the Function-as-a-Service (FaaS) ecosystem, including growth rates, architectural designs, and common use cases. Another study by (Eismann et al., 2021) examined different aspects such as implementation strategies, traffic patterns, and usage scenarios, focusing on serverless applications developed using a greenfield approach.

Additionally, (Leitner et al., 2019) conducted a mixed-method empirical study that underscored the necessity of adopting a different mental model when transitioning to serverless architectures and identified various prevalent application patterns. The research conducted by (Hamza et al., 2024) highlights that practitioners need to master several tools and technologies for the successful adoption of the serverless computing paradigm within their organizations.

Similarly, research in other software engineering domains has identified critical skills required for practitioners working with AI, requirements engineering, and testing. While automated methods can recognize the skills of software engineers, they often miss subtle nuances present in job advertisements (Papoutsoglou et al., 2017). For example, a study that analyzed 101 Dutch job openings in Requirements Engineering (RE) successfully categorized the essential required skills (Daneva et al., 2017). Moreover, a detailed examination of job advertisements has distinguished specific testing skills necessary for both coders and testers, emphasizing the importance of automated testing tools (Cerioli et al., 2020). The existing literature also thoroughly covers the desired profiles and competencies of AI engineers (Meesters et al., 2022) and similarly in the microservices architecture (Ayas et al., 2024).

However, current research does not effectively help practitioners prioritize emerging technologies or plan their professional development for serverless computing. Practitioners often lack specialized training, and the industry has not defined essential training for serverless application development. This has resulted in a significant research gap concerning the specific responsibilities and skills required for serverless computing.

## 3. Methodology

In this study, we employ qualitative analysis methods, specifically thematic analysis techniques (Braun & Clarke, 2006) to thoroughly examine the roles, responsibilities, and skills required for serverless engineers. The thematic analysis was conducted on a dataset of job ads sourced from Glassdoor, a leading global recruitment platform with millions of job listings. Further, the study is structured into four distinct phases as illustrated in Figure 1.

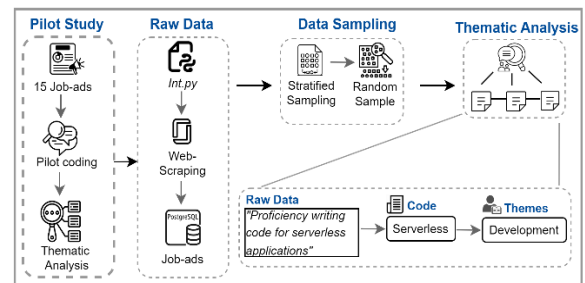

**Figure 1. Methodology.**

The first phase involved a pilot study, where the first author gathered fifteen job ads. The pilot samples were analyzed to identify potential insights related to required roles, responsibilities, and skills and to refine the research questions. Based on the findings of pilot samples, we proceeded with an extensive data collection phase, which resulted in the systematic compilation of 2,508 job advertisements. Subsequently, we applied stratified sampling to select a subset of the entire data and selected 141 job posts to qualitatively analyze.

## 3.1. Data collection

The primary goal was to systematically collect job ads featuring serverless computing skills globally. We developed an automated Python script using *Selenium*, *Selenium Stealth*, and *PostgreSQL* to retrieve real-time job listings from Glassdoor. For instance, the script extracted 800 U.S. job ads posted by April 5, 2024. The automation ensured data integrity by sorting listings chronologically and minimizing selection bias from Glassdoor's algorithms influenced by paid promotions and personalized user data.

The job posts were gathered from seven English-speaking countries across five continents: the United States, United Kingdom, Australia, United Arab Emirates, India, Singapore, and South Africa. This selection ensured English consistency and diverse representation, resulting in a dataset of 2,508 job ads. The script also collected metadata for each job ad, including recruiter details shown in Figure 2, enriching our analysis of regional and industry-specific demands for serverless skills.

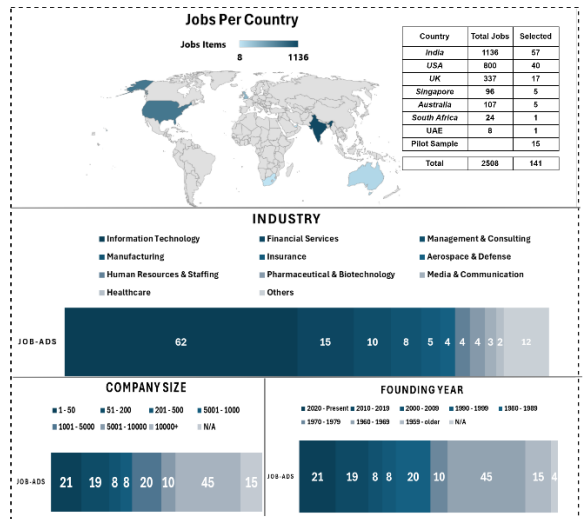

**Figure 2. Demographics**

## 3.2. Data sampling

After collecting job ads from Glassdoor, we used stratified random sampling for thematic analysis, ensuring representative selection across different regions (Baltes & Ralph, 2022). This method is widely used in software engineering, market research, and social sciences. It segments data by country, treating each as a separate stratum. Our dataset included 2,508 job ads from seven countries. For each stratum, we sampled 5% of the total dataset, calculating the sample size for each country using the formula $n_i = \lfloor 0.05 \times N_i \rfloor$. Here, $N_i$ represents the total number of jobs in the country $i$. This approach yielded the following sample sizes: US (40), UK (17), AUS (5), India (57), UAE (1), South Africa (1), and Singapore (5).

After implementing stratified sampling, we used simple random sampling with *Python's pandas library* to ensure each data point had an equal chance of selection (Lohr, 2021). This approach ensures our sample accurately reflects the larger dataset by maintaining statistical integrity. The replication package for the verification of our analysis is available online (Muhammad, n.d.).

## 3.3. Data analysis

This study employs a thematic analysis approach (Braun & Clarke, 2006) to analyze the data. This involved systematically coding and categorizing the data to identify significant themes. The process began with familiarization and reviewing the job ads to gain an initial understanding. We then generated initial codes, grouped them into sub-themes, and categorized them into broader main themes.

The first author analyzed 20 job ads, followed by the second and third authors each analyzing 20 ads. They continued this process until all 141 ads were analyzed. They then collaboratively finalized the sub-themes and main themes, ensuring consistent categorization. Finally, the fifth author an expert empirical researcher, reviewed and validated the themes, ensuring reliability and credibility.

## 4. Results

The analysis of 141 job posts provides insights into serverless practitioner roles, highlighting key responsibilities and required skills. We identified seven roles: *software engineer (44 ads), data engineer (27 ads), architect (21 ads), web developer (18 ads), cloud engineer (14 ads), engineering leadership and management (14 ads), and product manager (3 ads)*. We further identified nineteen responsibilities and grouped them into four themes. Additionally, we mapped the hard and soft skills required for each role, offering a comprehensive overview of key competencies required from serverless practitioners.

***RQ1: What are the roles and responsibilities of serverless practitioners as identified in job ads?***

We identified seven distinct roles for practitioners working with serverless computing, each with unique responsibilities. Furthermore, we identified 19 different responsibilities required by serverless practitioners, which we categorized into four overarching themes: *software development, infrastructure and operations, professional development and leadership, and software business*. Figure 3 illustrates the taxonomy of responsibilities.

***T1: Software Development***: in the context of serverless computing involves the systematic process of designing, creating, testing, and maintaining software applications. This theme is critical for ensuring that applications are not only functional but also secure, scalable, and efficient. The sub-themes within software development highlight various aspects of this process. *Architectural design* focuses on structuring software systems to align with business goals and technical requirements. *"Knowledge of architectural design patterns, performance tuning, database, and functional designs." – job-ad # 26*. *Secure software development* practices emphasize integrating security measures throughout the development lifecycle to safeguard applications from vulnerabilities. *"Partner with teams across the organization to perform architecture reviews, code security reviews, and promote secure development practices." - Job-ad # 28*. *Testing and quality assurance* ensure that software meets specified requirements and is free of defects, thus guaranteeing reliability and performance. *"Write unit and integration tests which will pave the way for continuous deployment and aim for zero bugs." – Job-ad # 61*. *Integration and API management* are vital for enabling seamless communication between different software components. *"Hands-on API and API Gateway implementations, managing integrations with external systems and internal modules/components." – Job-ad # 90*.

Apart from these themes, other sub-themes include, *Frontend development* ensures that the user interface is intuitive and user-friendly, while *Backend development* handles the server-side logic, databases, and APIs that power the front end. *Software process improvement* involves strategies to enhance development efficiency and effectiveness such as implementing agile software development. *Development tools and technologies* support these activities through IDEs, version control systems, and CI/CD pipelines. *Data management and analytics* ensure the integrity, accessibility, and actionable insights of data. Lastly, *AI/ML integration* incorporates AI capabilities to enhance application functionality with advanced features. Together, these sub-themes provide a comprehensive framework for understanding the diverse aspects of software development in serverless computing.

***T2: Infrastructure and Operations:*** While serverless computing abstracts away much of the underlying infrastructure, our findings reveal that companies still require serverless practitioners to possess a certain level of knowledge in managing the infrastructure. This insight underscores the importance of understanding and handling the foundational elements of serverless environments to ensure optimal performance, scalability, and reliability. This theme includes key sub-themes: *configuration and management of infrastructure*, which ensures scalable, reliable, and efficient infrastructure for serverless applications. *"Experience with Infrastructure as Code tools, particularly Terraform and Packer." – Job-ad # 136*. *DevOps automation tools*, which streamline development and operational processes through continuous integration and deployment (CI/CD). *"Drive the implementation and maintenance of CICD pipelines to automate software delivery processes." – Job-ad # 58*. *Operations and support*, which involve monitoring performance, managing incidents, and providing technical support. *"Proactively identify and address performance bottlenecks, ensuring optimal system performance." – Job-ad # 141*.

***T3: Professional Development and Leadership*** emphasize the importance of continuous growth and strategic direction. First, *Leadership and Strategic Planning* highlight the necessity for serverless practitioners to possess strong leadership capabilities and strategic planning skills, enabling them to guide teams, make informed decisions, and set a clear vision to navigate the complexities of serverless environments effectively. *"Lead and contribute to complex technical projects and initiatives that span multiple engineering teams." – job-ad # 2*. Second, *Professional Development Skills* focus on the ongoing personal and professional growth of serverless practitioners, emphasizing the importance of staying current with industry trends, acquiring new skills, and fostering an environment of continuous improvement. *"Stay up-to-date with industry best practices and emerging trends in software development, especially serverless technologies, and actively contribute to process improvements and quality assurance initiatives within the organization." – job-ads # 11*.

Together, these sub-themes underscore the critical role of leadership and professional development in ensuring that serverless practitioners can lead their teams effectively, plan strategically, and continuously enhance their skills to adapt to the evolving landscape of serverless computing.

***T4: Software business*** underscores the intersection of technology and business strategies in the realm of serverless computing. This theme is divided into four key sub-themes. *Business Development* focuses on identifying and pursuing new market opportunities, fostering partnerships, and driving growth. *"Conduct customer research, pilots, and business impact analysis to identify new business opportunities and customer requirements." – job-ad #17. Innovation and Solution Design* emphasize the creation of novel solutions and the application of innovative technologies to solve business problems. *"Monitor and research emerging innovation to determine functionality that may provide business value." – Job-ad # 85. Technology Advancement Promotion* involves advocating for and implementing cutting-edge technologies to maintain a competitive edge. *"Work cross-functionally with product, sales, and customer success to increase product usage and customer adoption." – Job-ad #44.* Lastly, the *Go-To-Market Strategy* covers the planning and execution of strategies to successfully launch and position products in the market. *"Drive market expansion efforts by identifying target segments and developing tailored go-to-market strategies." – Job-ad # 63.*

Together, these sub-themes highlight the importance of combining technical expertise with business acumen to drive the success of serverless computing initiatives.

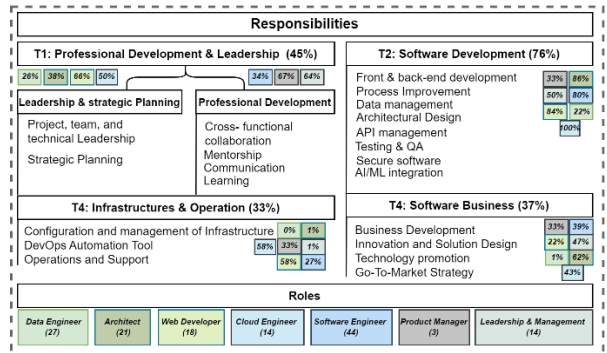

**Figure 3. Responsibilities**

*Responsibilities to roles*: Our analysis identified distinct distributions of responsibilities across various roles, calculated by the percentage of specific responsibilities per role. For instance, we examined how many of the 27 data engineer job ads require professional development and leadership skills. Figure 3 illustrates these findings. For instance, Software engineers are responsible for 80% of software development tasks, 39% of business activities, 34% of professional development, and 27% of infrastructure and operations. Data engineers contribute 22% to software development, and 26% to professional development, with minimal involvement in infrastructure and operations or business activities. Architects have 86% of their responsibilities in software development, 62% in business activities, and 38% in professional development, with little focus on infrastructure and operations. Web developers are tasked with 84% in software development, 58% in infrastructure and operations, 22% in business activities, and 66% in professional development. Cloud engineers are involved in 100% of software development, 58% of infrastructure and operations, 43% of business activities, and 50% of professional development. Further details on the roles are depicted in Figure 3.

*RQ2: What hard and soft skills are required from the serverless practitioner as identified in job ads?*

Beyond roles and responsibilities, our analysis identified 28 essential hard skills required for practitioners working with serverless computing. These skills range from software development, including frontend, backend, and cloud architecture, to software process improvement methodologies like Agile. We categorized these hard skills into seven main themes. *Software Development (Frontend, Backend, and Full Stack)*, *Software Operations*, *Cloud Architecture, Security and Compliance, Software Process Improvement, AI/ML, and Software Business*. Among them, three main themes for the hard skills are *cloud architecture and infrastructure, software development, and software operations.* These themes provide a comprehensive overview of the technical competencies needed to excel in serverless computing, highlighting the diverse and specialized skill sets required in this field are shown in Figure 4.

The **Cloud Architecture and Infrastructure** theme highlights the foundational elements essential for building and maintaining cloud-based systems. This theme includes several key sub-themes. *Serverless Computing* such as *AWS Lambda* focuses on the deployment and management of applications without the need for server management, promoting scalability and efficiency. *Big Data Management* emphasizes handling vast amounts of data, ensuring its storage, processing, and analysis are optimized for performance and reliability. *Scalable Systems and Distributed Computing* address the design and implementation of systems that can efficiently scale and operate across multiple distributed environments. *Application Migration* involves transitioning existing applications to cloud environments, ensuring minimal disruption and improved performance. *Architecture* covers the strategic design of cloud systems, ensuring they meet business and technical requirements. Lastly, *Software Design* focuses on the principles and practices of designing robust, efficient, and maintainable software systems within the cloud infrastructure. Together, these sub-themes provide a comprehensive framework for understanding the critical aspects of cloud architecture and infrastructure, highlighting the skills and knowledge required to excel in this domain. *"Work with Solution Architecture to influence solutions and ensure High-Level Designs are implemented as intended."* – *Job-ad # 76*

The **Software Development** theme encompasses essential skills for serverless computing. Key sub-themes include frontend and visualization development, backend development, and full-stack expertise. API development and integration, proficiency in programming languages, and knowledge of frameworks and operating systems are also critical. Additionally, microservices architecture, messaging queue implementation, rigorous testing, and quality assurance are important components. These sub-themes highlight the diverse and comprehensive skill set required for effective software development in serverless computing. *"Confidence to work in multiple programming languages could include C, Go, Rust, Python, Lua, and even PHP."* - *Job-ad # 69*

The **DevOps and Software Operations** theme is essential for serverless computing, focusing on the continuous integration and delivery (CI/CD) pipeline, DevOps practices, and version control. Key areas include software deployment, scripting automation, and infrastructure as code (IaC). Performance and observability ensure applications run smoothly and efficiently. These sub-themes highlight the critical skills required for maintaining robust and efficient serverless operations, ensuring seamless development, deployment, and monitoring of applications. *"Experience with IaC and Serverless tools like Terraform, Ansible, AWS Lambda."* -*Job-ad # 83*.

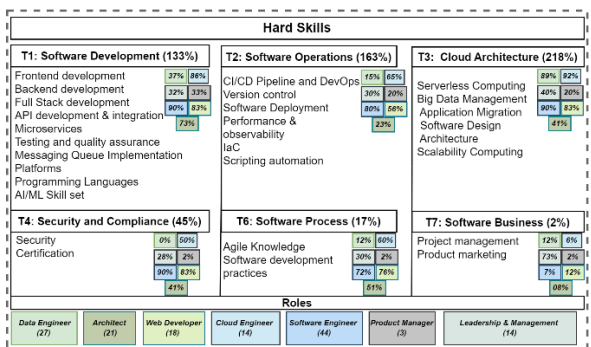

**Figure 4. Hard skills**

*Hard skills mapping to Roles:* Our analysis identified distinct distributions of hard skills across various roles, calculated by the percentage of specific skills per role. For instance, we examined how many of the 27 data engineer job ads require skills in cloud architecture. Data engineers are responsible for 37% of software development tasks, while cloud engineers, web developers, and architects have higher involvement at 86%, 83%, and 73% respectively. Engineering leadership and management roles allocate 32% of their responsibilities to software development, while product managers have a lower engagement at 33%. Software engineers have the second highest involvement, with 90% of their responsibilities related to software development. In software operations, data engineers and architects are involved in 15% and 23% of tasks, whereas cloud engineers and web developers have higher involvement at 65% and 56%. Engineering leadership and management roles contribute 30% of their responsibilities to software operations, and product managers are engaged at 20%. Software engineers have the highest involvement at 80%. For cloud architecture, data engineers are significantly involved with 89% of their tasks related to this area. Cloud engineers and architects have even higher involvement at 92% and 86%. Engineering leadership and management roles dedicate 40% of their responsibilities to cloud architecture, while product managers are engaged at 20%. Software engineers and web developers also have substantial involvement, with 90% and 83% of their responsibilities related to cloud architecture. Regarding software process improvement, data engineers are involved in 12% of these tasks. Cloud engineers have more significant involvement at 60%. Engineering leadership and management roles dedicate 30% of their responsibilities to process improvement, while product managers have minimal involvement at 2%. Software engineers, architects, and web developers are highly engaged, with 72%, 51%, and 76% of their responsibilities respectively related to software process improvement. Figure 4 presents the complete taxonomy.

*Soft Skills:* based on the analysis of 141 job posts, we identified 32 sub-themes that were categorized into 7 main themes representing the soft skills required for serverless practitioners. Among these, the six most demanded soft skills across all job ads are (i) communication proficiency, (ii) continuous learning and adaptability, (iii) collaborative teamwork, (iv) problem-solving and analytical skills, (v) leadership excellence, and (vi) project management. We further mapped the soft skills to the identified roles. The thematic findings of this research question are depicted in Figure 5.

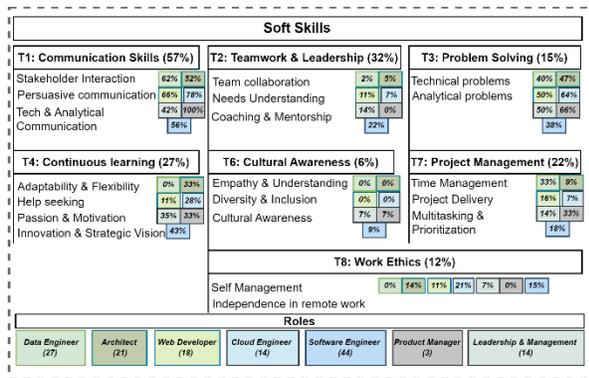

Figure 5. Soft skills

*Communication proficiency* is essential for clear and effective information exchange, enabling seamless collaboration and understanding among team members, stakeholders, and clients. This skill ensures that ideas, feedback, and instructions are conveyed accurately, reducing the risk of misunderstandings and errors. An example of the required soft skill in communication is: "*Strong ability to communicate with a broad range of clients, colleagues, and partners across a variety of contexts and formats.*" – Job-ad # 15. *Continuous learning and adaptability* are essential for staying ahead in the rapidly evolving field of serverless architecture. These skills enable practitioners to stay current and proficient, seamlessly integrating new tools and techniques into their workflows and adapting to changing project requirements. An example of the required soft skill in continuous learning and adaptability is: "*Stay up-to-date, constantly learning about advances in the field, and deliver periodic presentations to internal teams on these developments.*" – Job-ads # 45. *Collaborative teamwork* is vital for successful project integration, fostering a cooperative environment where diverse skills and perspectives contribute to common goals. Effective teamwork enhances problem-solving capabilities, accelerates project timelines, and improves overall project quality by leveraging the team's collective expertise. An example of collaborative teamwork is: "*Collaborate with cross-functional teams including product managers, QA engineers, and designers to ensure successful delivery and alignment with business requirements.*" – Job-ad 1. *Problem-solving and analytical skills* are necessary for identifying and addressing technical challenges, allowing practitioners to devise innovative and efficient solutions. "*Solves problems that have unique and/or broad implications for the technology architecture.*" – Job-ad # 43. *Leadership excellence* is important for guiding and inspiring teams, providing direction and motivation to achieve project objectives, and drive organizational success. Leaders in serverless architecture set the vision, allocate resources, and cultivate a positive and productive team culture, enabling high performance and successful project outcomes. An example of leadership excellence soft skill is. "*Ensure high-quality deliverables by demonstrating leadership in code reviews and test practices.*" – Job-ad 48. Finally, *project management* is key for ensuring project goals and timelines are met and coordinating resources and tasks to deliver successful outcomes within set parameters. An example of the project management soft skill is. "*Manage sole project priorities, deadlines, and deliverables.*" – Job-ad # 99

The soft skills required for a *data engineering role* are not evenly distributed across all the main themes. However, key skills include communication proficiency (62%), problem-solving and analytical skills (41%), and project management skills (33%). However, no other soft skills are required for the data engineering role. These skills are crucial for effectively performing the role and contributing to successful project outcomes. Similarly, the *architect* role emphasizes communication proficiency (52%), problem-solving and analytical skills (47%), and continuous learning and adaptability (34%). However, empathy and cultural awareness soft skill is not mentioned in any of the job posts for the architect role. These soft skills are vital for excelling in the role and ensuring effective project execution and adaptation to new challenges. For *web developers*, communication proficiency (66%) and project management skills (16%) are crucial. However, skills such as sales and customer orientation, work ethic, and independence are not commonly mentioned in job descriptions for web developers. For *cloud engineers*, the essential soft skills include communication proficiency (78%), problem-solving and analytical skills (64%), and continuous learning and adaptability (28%). These skills are critical for navigating the dynamic landscape of cloud technology, enabling engineers to effectively collaborate, tackle complex challenges, and stay ahead of industry advancements. For the *engineering leadership and management role*, all the identified soft skills were required in almost all the job ads related to engineering leadership and management role. For instance, communication skills (42%), continuous learning and adaptability (35%), and Teamwork and leadership excellence (14%) are particularly in demand. These skills are vital for effectively guiding teams, fostering innovation, and maintaining a competitive edge in the industry. For the *product manager role*, essential soft skills include communication proficiency (100%), problem-solving and analytical skills (66%), continuous learning and adaptability (33%), and project management skills (33%). Finally, for the *software engineer role*, the most essential skills mentioned in the job ads are communication skills (56%), problem-solving and analytical skills (38%), and continuous learning and adaptability (43%).

> **Takeaway 1:** We found that communication skills, problem-solving and analytical skills, and continuous learning and adaptability are the most in-demand soft skills across all identified roles.

*RQ3: What are the trends in programming languages, serverless technologies, and cloud platforms as indicated by serverless- job ads?*

We investigated evolving trends in serverless computing by analyzing job ads for programming languages, serverless technologies, and cloud platforms. This analysis identifies the most in-demand programming languages, popular serverless frameworks, and platforms, aiming to inform future educational and training programs.

*Programming Languages:* Programming language preferences in serverless job ads reveal an evolving software development landscape. Our analysis shows Python leading with 32 mentions (22.7%), highlighting its widespread adoption and versatility. Java and JavaScript follow with 27 (19.15%) and 23 (16.6%) mentions, respectively, reflecting their importance in enterprise applications and web services. Emerging languages like Node.js and TypeScript also show significant traction with 22 (15.6%) and 9 (6%) mentions, indicating a shift towards JavaScript-based technologies for better scalability and maintainability. Other languages such as C#, .Net, and Go are also noted in our replication package (Muhammad, n.d.).

We further analyzed programming language trends in relation to specific roles and found *Java* and *JavaScript* are the most demanded for software engineering. *Node.js* is trending for engineering leadership and management roles. *Python* is highly sought after in data engineering and architect roles. For web developers, *JavaScript* and *Node.js* are the most demanded programming languages. Similarly, *Node.js* and *Python* are the preferred languages for cloud engineer roles. The results are depicted in Figure 6.

*Serverless Technologies:* The analysis of serverless technologies in job advertisements reveals a significant dominance of *AWS Lambda* with 62 (45%) mentions, underscoring its preeminence in serverless computing. *Azure Functions*, while significantly trailing with 12 (9%) mentions, indicates Microsoft's growing influence in the market. The data also highlights a robust integration of AWS services, including API Gateway 11 (8%), SNS 7 (5%), SQS 7 (5%), DynamoDB 5 (4%), and other 34 (24%) reflects a preference for AWS's extensive suite of tools that cater to various aspects of serverless computing. This trend suggests that professionals in the serverless domain should focus on acquiring expertise in AWS technologies, while also considering the emerging relevance of Azure Functions to remain competitive in the job market.

*Platforms:* The analysis of cloud platform mentions in job advertisements shows *AWS* as the predominant platform with 83 mentions (58%), indicating its strong leadership in the serverless computing market. *Azure* follows with 36 mentions (25%), demonstrating its significant presence as a competitive alternative. *Google Cloud Platform (GCP)* has 22 mentions (17%), highlighting its niche but growing role in the serverless ecosystem. However, we did not find any open-source serverless frameworks such as OpenFaas, or Kubeless mentioned in any job posts, indicating a low trend in choosing open-source solutions for developing serverless applications.

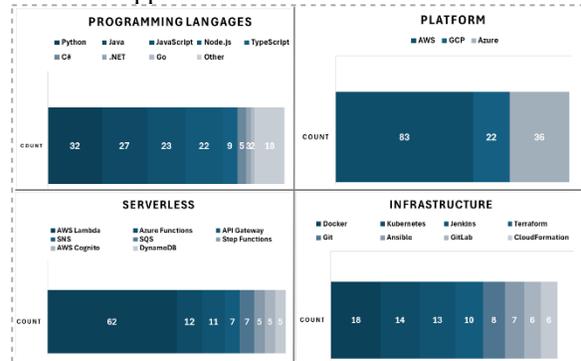
**Figure 6. Technologies implementation**

Therefore, to excel in serverless computing, practitioners should focus on mastering Python, Java, JavaScript, Node.js, and TypeScript, along with key AWS services like Lambda, API Gateway, SNS, SQS, and DynamoDB. Additionally, proficiency in Azure Functions and Google Cloud Platform will provide a competitive edge.

## 5. Summary

This study offers a comprehensive understanding of the roles, responsibilities, and skills essential for serverless practitioners, providing valuable insights for practitioners, educators, and organizations seeking to align with industry standards. To achieve this, we extracted 2,508 job ads from Glassdoor across seven English-speaking countries and selected 141 job ads through stratified sampling. These selected job ads were then analyzed using thematic analysis.

*Organizational Structure:* our analysis conducted on the job ads revealed that organizations adopting serverless computing must address skill gaps by investing in continuous learning and professional development. Companies should prioritize hiring individuals with strong communication, problem-solving, and leadership skills to navigate the complexities of serverless projects successfully.

***Implications for Education and Training:*** Educational and training programs should equip practitioners with a blend of both hard and soft skills necessary for success in serverless computing. These programs should emphasize cloud architecture, software development, and process improvement while integrating training in communication, continuous learning, problem-solving, leadership, and project management. By doing so, training programs can ensure that practitioners are well-prepared to handle the technical and collaborative demands of the industry.

***Clarifying the Roles and Responsibilities:*** The analysis confirmed key roles in serverless computing, including software engineers, data engineers, architects, web developers, cloud engineers, engineering leadership, and product managers. Each role encompasses distinct responsibilities, with software engineers and cloud engineers heavily involved in both development and operations, highlighting the need for a comprehensive skill set.

***Soft Skills as a Critical Component:*** our analysis conducted on the job ads reveals that communication proficiency, continuous learning, and problem-solving are paramount across all roles. These skills are essential for effective collaboration, adaptation to new technologies, and innovative problem-solving. Leadership and project management skills are particularly vital for roles involving team guidance and strategic oversight, such as engineering leadership and management. This indicates the importance of balancing technical expertise with strong interpersonal and organizational skills to succeed in serverless environments.

## 6.1. Threats to Validity

This study acknowledges several threats to validity that could impact the findings. These threats are categorized into external, internal, and construct.

**External validity** is the extent to which study findings can be applied beyond the sample and context. Our data, gathered exclusively from Glassdoor, limits the representativeness of serverless engineering roles, responsibilities, and skills across the software industry, as not all employers use this platform. To mitigate this, we included data from a broad geographic range, focusing on countries with a strong Glassdoor presence (Yin, 2018). Additionally, some companies do not fully disclose roles and skills in job descriptions, limiting generalizability. To address this, we included all job posts mentioning 'serverless' in their descriptions.

**Internal validity** refers to the extent to which study results can be attributed to the variables under investigation rather than other factors. Job ads reflect current requirements, which may evolve or differ from actual responsibilities. To mitigate this, we focused on frequently mentioned responsibilities and skills (Easterbrook et al., 2008) and used stratified sampling for a representative sample. Despite potential selection bias from underrepresented subgroups in the Glassdoor dataset, we ensured our sampling method accurately reflected job ads' distribution. We also cross-validated our sample with additional job ad sources to confirm representativeness.

**Construct validity** refers to the extent to which our measurements accurately reflect the concepts being studied. The thematic analysis approach relies on consistent coding and categorization of data. Variations in the interpretation of job descriptions and the subsequent coding process can threaten construct validity (Trochim et al., 2016). To mitigate this threat, all the authors conducted rigorous thematic analysis, and we collaboratively developed distinct, well-defined themes to enhance the validity of our findings.

## 7. Conclusion and future work

As serverless architecture gains traction in the industry, the underlying technologies are becoming increasingly diverse. This study develops a comprehensive taxonomy of the roles, responsibilities, and skills required from serverless practitioners, based on the analysis of 141 job ads from Glassdoor. We identified 19 distinct responsibilities, categorized into four overarching themes: *software development, infrastructure and operations, professional development and leadership, and software business*. Additionally, we identified 28 hard skills, grouped into seven themes: *Software Development (Frontend, Backend, and Full Stack), Software Operations, Cloud Architecture, Security and Compliance, Software Process Improvement, AI/ML, and Software Business*. We also identified 32 soft skills, mapped into 10 themes, with the six most demanded being *communication proficiency, continuous learning and adaptability, collaborative teamwork, problem-solving and analytical skills, leadership excellence, and project management*. Finally, we discussed directions for future research, focusing on unraveling organizational structures and personal development for practitioners in serverless computing.

Future research should aim to expand the scope of this study by including a broader range of data sources beyond Glassdoor to capture a more comprehensive view of the global job market for serverless practitioners. Longitudinal studies could also provide deeper insights into the evolving trends and changing demands in serverless computing over time.

**AI disclosure:** We disclose that we utilized large language models to enhance the language and structure of the research paper.

## 8. References

Ali, M., Ali, S., & Jilani, A. (2020). Architecture for microservice based system. A report. A Report. https://www.researchgate.net/profile/Muzaffar-Temoor/publication/348869792_Architecture_for_Microservice_Based_System_A_Report/links/6013ca43a6fdcc071b9d2c35/Architecture-for-Microservice-Based-System-A-Report.pdf

Ayas, H. M., Hebig, R., & Leitner, P. (2024). An empirical investigation on the competences and roles of practitioners in Microservices-based Architectures. Journal of Systems and Software, 213, 112055.

Baldini, I., Castro, P., Chang, K., Cheng, P., Fink, S., Ishakian, V., Mitchell, N., Muthusamy, V., Rabbah, R., Slominski, A., & Suter, P. (2017a). Serverless Computing: Current Trends and Open Problems. In S. Chaudhary, G. Somani, & R. Buyya (Eds.), Research Advances in Cloud Computing (pp. 1–20). Springer Singapore. https://doi.org/10.1007/978-981-10-5026-8_1

Baldini, I., Castro, P., Chang, K., Cheng, P., Fink, S., Ishakian, V., Mitchell, N., Muthusamy, V., Rabbah, R., Slominski, A., & Suter, P. (2017b). Serverless Computing: Current Trends and Open Problems. In S. Chaudhary, G. Somani, & R. Buyya (Eds.), Research Advances in Cloud Computing (pp. 1–20). Springer Singapore. https://doi.org/10.1007/978-981-10-5026-8_1

Baltes, S., & Ralph, P. (2022). Sampling in software engineering research: A critical review and guidelines. Empirical Software Engineering, 27(4), 94. https://doi.org/10.1007/s10664-021-10072-8

Braun, V., & Clarke, V. (2006). Using thematic analysis in psychology. Qualitative Research in Psychology, 3(2), 77–101. https://doi.org/10.1191/1478088706qp063oa

Brooks, F. C. (2021). The Mythical Man-Month (1975). https://direct.mit.edu/books/edited-volume/chapter-pdf/2248398/9780262363174_c003900.pdf

Cerioli, M., Leotta, M., & Ricca, F. (2020). What 5 million job advertisements tell us about testing: A preliminary empirical investigation. Proceedings of the 35th Annual ACM Symposium on Applied Computing, 1586–1594. https://doi.org/10.1145/3341105.3373961

Daneva, M., Wang, C., & Hoener, P. (2017). What the job market wants from requirements engineers? An empirical analysis of online job ads from the Netherlands. 2017 ACM/IEEE International Symposium on Empirical Software Engineering and Measurement (ESEM), 448–453. https://ieeexplore.ieee.org/abstract/document/8170133/

Easterbrook, S., Singer, J., Storey, M.-A., & Damian, D. (2008). Selecting Empirical Methods for Software Engineering Research. In F. Shull, J. Singer, & D. I. K. Sjøberg (Eds.), Guide to Advanced Empirical Software Engineering (pp. 285–311). Springer London. https://doi.org/10.1007/978-1-84800-044-5_11

Eismann, S., Scheuner, J., Van Eyk, E., Schwinger, M., Grohmann, J., Herbst, N., Abad, C. L., & Iosup, A. (2021). The state of serverless applications: Collection, characterization, and community consensus. IEEE Transactions on Software Engineering, 48(10), 4152–4166.

Eivy, A., & Weinman, J. (2017). Be wary of the economics of" serverless" cloud computing. IEEE Cloud Computing, 4(2), 6–12.

Eskandani, N., & Salvaneschi, G. (2023). The uphill journey of FaaS in the open-source community. Journal of Systems and Software, 198, 111589.

Hamza, M. (2023). Software Architecture Design of a Serverless System. Proceedings of the 27th International Conference on Evaluation and Assessment in Software Engineering, 304–306. https://doi.org/10.1145/3593434.3593471

Hamza, M., Akbar, M. A., & Smolander, K. (2024). The Journey to Serverless Migration: An Empirical Analysis of Intentions, Strategies, and Challenges. In R. Kadgien, A. Jedlitschka, A. Janes, V. Lenarduzzi, & X. Li (Eds.), Product-Focused Software Process Improvement (Vol. 14483, pp. 100–115). Springer Nature Switzerland. https://doi.org/10.1007/978-3-031-49266-2_7

Leitner, P., Wittern, E., Spillner, J., & Hummer, W. (2019). A mixed-method empirical study of Function-as-a-Service software development in industrial practice. Journal of Systems and Software, 149, 340–359.

Lohr, S. L. (2021). Sampling: Design and analysis. Chapman and Hall/CRC. https://www.taylorfrancis.com/books/mono/10.1201/9780429298899/sampling-sharon-lohr

Meesters, M., Heck, P., & Serebrenik, A. (2022). What is an AI engineer?: An empirical analysis of job ads in The Netherlands. Proceedings of the 1st International Conference on AI Engineering: Software Engineering for AI, 136–144. https://doi.org/10.1145/3522664.3528594

Montandon, J. E., Politowski, C., Silva, L. L., Valente, M. T., Petrillo, F., & Guéhéneuc, Y.-G. (2021). What skills do IT companies look for in new developers? A study with Stack Overflow jobs. Information and Software Technology, 129, 106429.

Muhammad, H. (n.d.). Unveiling the Skills and Responsibilities of Serverless Practitioners: An Empirical Investigation [dataset]. https://doi.org/10.5281/zenodo.11622664

Papoutsoglou, M., Mittas, N., & Angelis, L. (2017). Mining people analytics from stackoverflow job advertisements. 2017 43rd Euromicro Conference on Software Engineering and Advanced Applications (Seaa), 108–115. https://ieeexplore.ieee.org/abstract/document/8051336/

Trochim, W. M., Donnelly, J. P., & Arora, K. (2016). Research methods: The essential knowledge base. Cengage learning. https://iro.uiowa.edu/esploro/outputs/9984214724402771?institution=01IOWA_INST&skipUsageReporting=true&recordUsage=false

Wen, J., Chen, Z., Liu, Y., Lou, Y., Ma, Y., Huang, G., Jin, X., & Liu, X. (2021). An empirical study on challenges of application development in serverless computing. Proceedings of the 29th ACM Joint Meeting on European Software Engineering Conference and Symposium on the Foundations of Software Engineering, 416–428. https://doi.org/10.1145/3468264.3468558

Yin, R. K. (2018). Case study research and applications (Vol. 6). Sage Thousand Oaks, CA. https://www.academia.edu/download/106905310/Artikel_Yustinus_Calvin_Gai_Mali.pdf